\documentclass[preprint,showpacs,preprintnumbers,amsmath,amssymb]{revtex4}

\usepackage{natbib}
\usepackage{graphicx}
\usepackage{hyperref}
\usepackage{amsmath}
\usepackage{amssymb}
\usepackage{amsfonts}

\usepackage[T3,T1]{fontenc}
\DeclareSymbolFont{tipa}{T3}{cmr}{m}{n}
\DeclareMathAccent{\invbreve}{\mathalpha}{tipa}{16}

\begin{document}

\renewcommand{\figureautorefname}{Fig.}
\renewcommand{\equationautorefname}{Eq.}

\vspace*{1cm}
\begin{center}
{\LARGE{\textbf{Correcting for imperfectly sampled data in the iterative Marchenko scheme}}} \\
\vspace{1cm}
Johno van IJsseldijk and Kees Wapenaar
\end{center}

\textbf{Summary}

The Marchenko method retrieves the responses to virtual sources in the subsurface, accounting for all orders of multiples. The method is based on two integral representations for focusing and Green's functions. In discretized form these integrals are represented by finite summations over the acquisition geometry. Consequently, the method requires ideal geometries of regularly sampled and co-located sources and receivers. However, a recent study showed that this restriction can, in theory, be relaxed by deconvolving the irregularly-sampled results with certain point spread functions (PSFs).The results are then reconstructed as if they were acquired using a perfect geometry. Here, the iterative Marchenko scheme is adapted in order to include these PSFs; thus, showing how imperfect sampling can be accounted for in practical situations. Next, the new methodology is tested on a 2D numerical example. The results show clear improvement between the proposed scheme and the standard iterative scheme. By removing the requirement for perfect geometries the Marchenko method can be more widely applied to field data.

\newpage

\section{Introduction}

Deep seismic targets are often obstructed by shallower structures in the subsurface. These shallow reflections and their multiples should ideally be removed from the data to retrieve a better image from the deep target. This can be achieved by virtually moving the wavefield recorded at the surface to a new acquisition level in the subsurface using the Marchenko method. This data-driven method redatums sources and receivers as if they were located at an arbitrary position inside the subsurface, while accounting for all orders of multiples \citep{broggini2012focusing,wapenaar2014marchenko,slob2014seismic}. \\
Recent applications of the Marchenko method to field data show great potential \citep[e.g.][]{ravasietal2016,staringetal2018,Brackenhoff2019}. However, strict requirements on the acquisition geometry obstruct more wide-spread use of the method. Non-perfect geometries can significantly distort the results \citep{peng2019effects,staring2019interbed}. Therefore, most authors tacitly assume regularly sampled and collocated sources and receivers. Ideally, this restriction should be relaxed or even removed, allowing for broader application of the method on field data. \\ 
Irregular sampling over the receiver dimension can be corrected for using sparse inversion \citep{ravasi2017rayleigh,haindl2018sparsity}, while irregular sampling over the source dimensions can, in theory, be corrected for using point-spread functions \citep[PSFs,][]{wapenaar2019}. Here, we explore how these PSFs can be integrated into the iterative Marchenko scheme \citep{thorbecke2017implementation} in order to handle irregular source sampling in practical applications.

\section{Theory}

\newcommand{\xs}[1]{\textbf{x}_{#1}}
Imagine an inhomogeneous lossless subsurface bounded by transparent acquisition surface $\mathbb{S}_0$. The reflection response at this surface is given by $R(\xs{R},\xs{S},\omega)$, where $\xs{R}$ and $\xs{S}$ are the receiver and source positions, respectively, and $\omega$ denotes the angular frequency. We define the virtual acquisition depth at $\mathbb{S}_A$, on which the virtual receivers are located. These receivers are used to measure the up- and down-going Green's functions: $G^-(\xs{A},\xs{R},\omega)$ and $G^+(\xs{A},\xs{R},\omega)$, respectively. Here, $\xs{A}$ is the location of the virtual receivers at the virtual acquisition depth. Next, the medium is truncated below $\mathbb{S}_A$, resulting in a medium that is inhomogeneous between $\mathbb{S}_0$ and $\mathbb{S}_A$, and homogeneous above and below these surfaces. In this medium we define a downgoing focusing function $f^+_1(\xs{S},\xs{A},\omega)$, which, when injected from the surface, focuses at the focal depth $\mathbb{S}_A$. Moreover, $f^-_1(\xs{R},\xs{A},\omega)$ is the upgoing response of the medium as measured on the surface, known as the upgoing focusing function. These ideas can be combined in two integral equations, as follows \citep{wapenaar2014marchenko,slob2014seismic}:
\begin{equation}
\label{eqn:mar1}
G^-(\xs{A},\xs{R},\omega) + f^-_1(\xs{R},\xs{A},\omega) = \int_{\mathbb{S}_0} R(\xs{R},\xs{S},\omega) f^+_1(\xs{S},\xs{A},\omega)d\xs{S} \text{,}
\end{equation}
\begin{equation}
 \label{eqn:mar2}
G^+(\xs{A},\xs{R},\omega) - \{f^+_1(\xs{R},\xs{A},\omega)\}^* = -\int_{\mathbb{S}_0} R(\xs{R},\xs{S},\omega) \{f^-_1(\xs{S},\xs{A},\omega)\}^*d\xs{S} \text{.}
\end{equation}
The asterisk $^*$ denotes complex conjugation. For acoustic media, the focusing and Green's functions on the left-hand side are separable in time by a windowing function. In practice, the infinite integrals on the right-hand side are  approximated by a finite sum over the available sources. When the reflection response is not well sampled, these summations cause distortions in the responses on the left-hand sides of \autoref{eqn:mar1} and \ref{eqn:mar2}. \\
\citet{wapenaar2019} introduce point-spread functions (PSFs) to correct for imperfect sampling. These PSFs exploit the fact that the downgoing focusing function is the inverse of the transmission response. A convolution of the focusing function with the transmission response should, therefore, give a delta pulse. However, for imperfectly sampled data this delta pulse gets blurred. This blurring caused by the imperfect sampling is quantified as follows:
\begin{equation}
\Gamma^+(\xs{A}',\xs{A},\omega)= \sum_{i} T(\xs{A}',\xs{S}^{(i)},\omega) f^+_1(\xs{S}^{(i)},\xs{A},\omega) S(\omega) \text{,}
\end{equation}
where $\Gamma^+$ and $T$ are the PSF and transmission response, respectively. Similarly, a quantity $Y$ is defined as the inverse of the conjugated, upgoing focusing function. Note that this inverse is not necessarily stable. Again, the convolution of $\{f^-_1\}^*$ with $Y$ results in a delta pulse for well-sampled data, but gets blurred for imperfectly sampled data. The second PSF ($\Gamma^-$) then becomes:
\begin{equation}
\Gamma^-(\xs{A}',\xs{A},\omega)= \sum_{i} Y(\xs{A}',\xs{S}^{(i)},\omega) \{f^-_1(\xs{S}^{(i)},\xs{A},\omega)\}^*S(\omega) \text{.}
\end{equation}
Next, these newly acquired PSFs are applied to \autoref{eqn:mar1} and \ref{eqn:mar2}, respectively:
\begin{equation}
\label{eqn:mar3}
\invbreve{G}^-(\xs{A},\xs{R},\omega) + \invbreve{f}^-_1(\xs{R},\xs{A},\omega) = \sum_{i} R(\xs{R},\xs{S}^{(i)},\omega)  f^+_1(\xs{S}^{(i)},\xs{A},\omega) S(\omega) \text{,}
\end{equation}
\begin{equation}
\label{eqn:mar4}
\invbreve{G}^+(\xs{A},\xs{R},\omega) - \{\invbreve{f}^+_1(\xs{R},\xs{A},\omega)\}^* = 
-\sum_{i} R(\xs{R},\xs{S}^{(i)},\omega) \{ f^-_1(\xs{S}^{(i)},\xs{A},\omega)\}^*S(\omega) \text{.}
\end{equation}
These equations have two interesting features. First, the right-hand sides now contain the desired summations. Second, the responses on the left-hand side now contain the PSFs, which apply a blurring effect to each response. In \autoref{eqn:mar3} the hat denotes a convolution with the downgoing PSF ($\Gamma^+$), whereas in \autoref{eqn:mar4} the responses with a hat are convolved with the upgoing PSF ($\Gamma^-$). Note that the imperfectly sampled Green's and focusing functions can now be deblurred by a multidimensional deconvolution (MDD) using the PSFs.
\begin{figure}[t!]
\centering
\includegraphics[bb=0 0 100 100,trim={0cm -1cm 0cm 0cm},clip,width=0.5\textwidth]{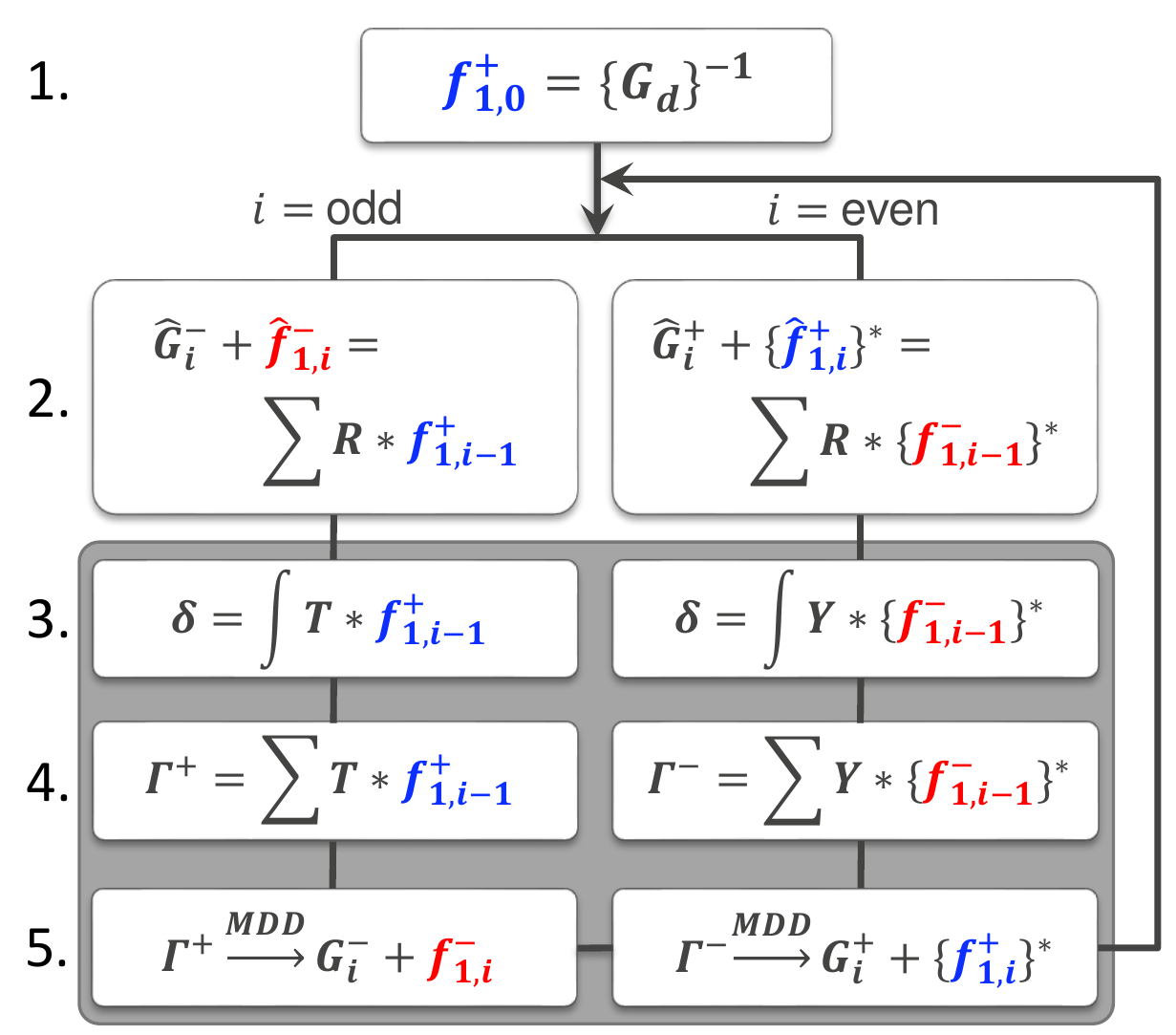}
\includegraphics[bb=0 0 325 325,width=0.48\textwidth]{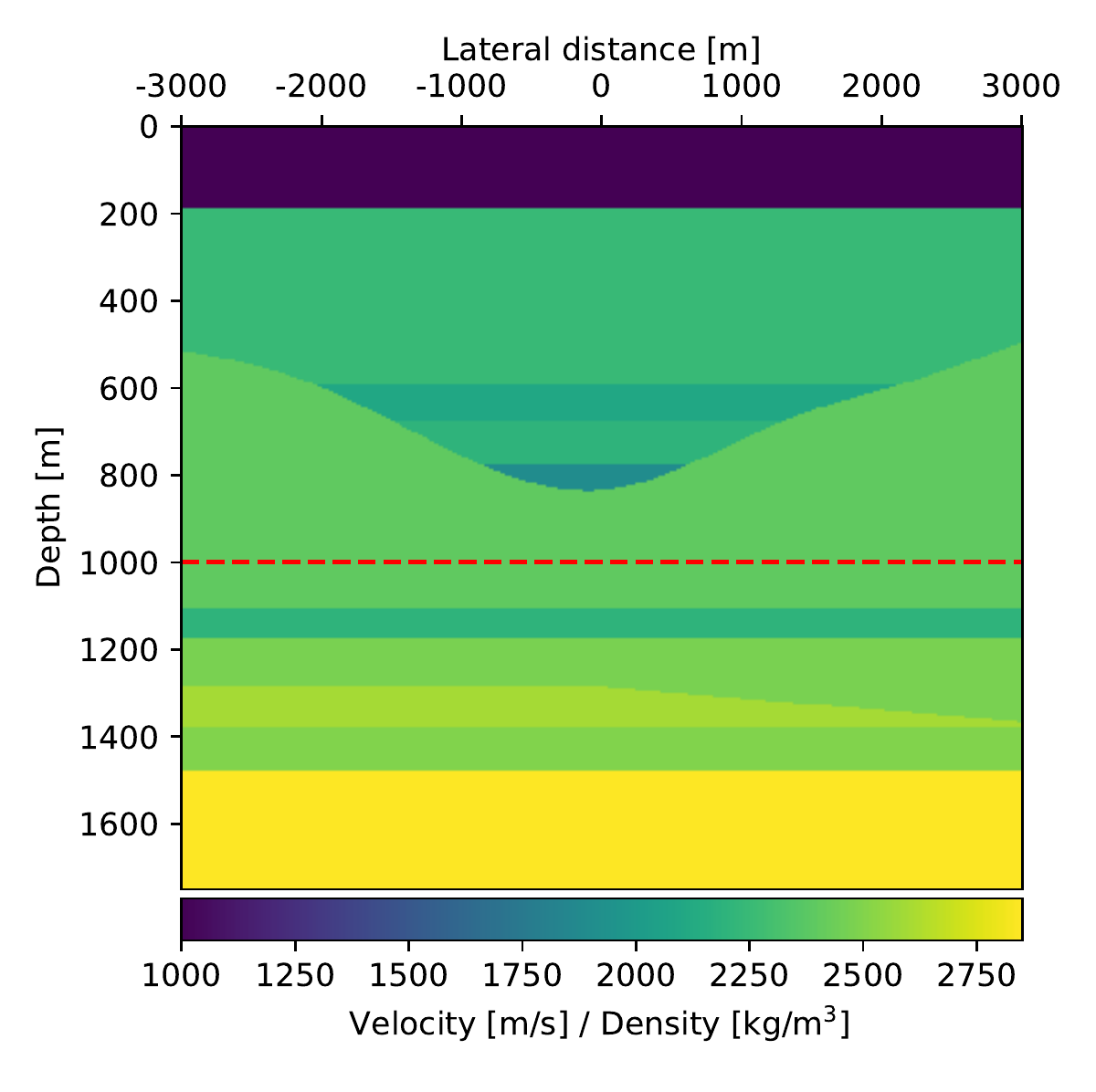} 
\caption{On the left, the flowchart of the proposed iterative Marchenko scheme, steps 3 to 5 account for imperfectly sampled data. The summation and integral denote convolution or correlation over the imperfectly sampled and well-sampled data, respectively. The right panel shows the velocity and density models for the numerical experiment, the virtual acquisition level is illustrated with the dashed red line.}
\label{is_model}
\end{figure} \\
However, since the PSFs are not known beforehand, their estimation will be incorporated into the iterative Marchenko scheme \citep{thorbecke2017implementation}, as shown in \autoref{is_model}. The first 2 steps are similar to the old scheme, with the only difference being that the inverse of the direct Green's function is used as opposed to a time-reversed version. Steps 3 to 5 are then introduced to reconstruct well-sampled responses from their blurred versions retrieved in step 2. First, the transmission response $T$ or quantity $Y$ are approximated, by inverting the focusing functions. Second, the PSFs are computed using these approximations and the irregular sampling of the sources. Finally, the PSFs are used to reconstruct the responses in step 2 as if they were regularly sampled. These deblurred responses can be separated in time, just like the standard Marchenko method. Each iteration of the scheme then starts with a deblurred response, which is computed in the previous iteration.


\section{Results}


\begin{figure}[b!]
\centering

\includegraphics[width=\textwidth]{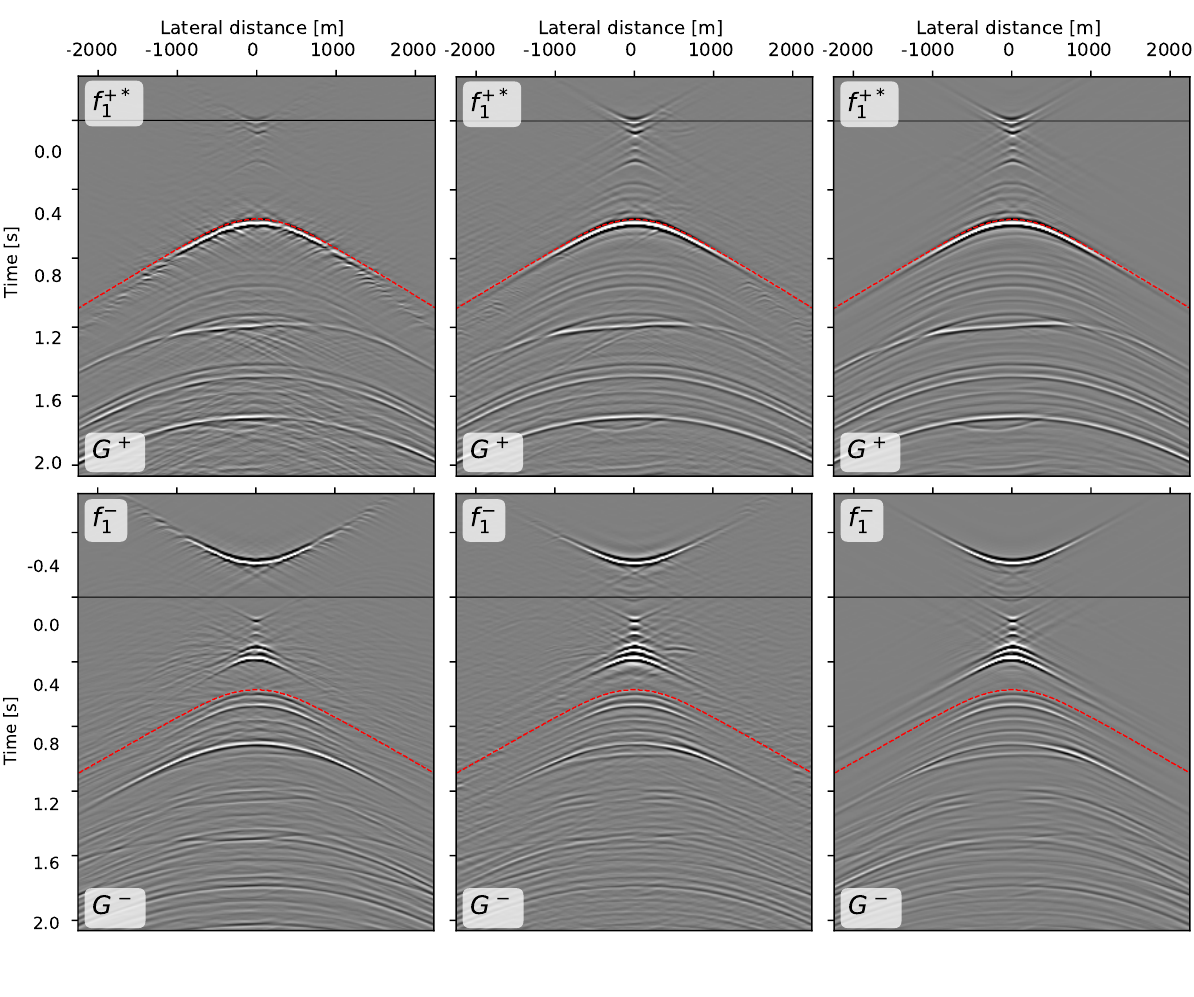}
\caption{The top row shows the time-reversed downgoing focusing function ($\{f_1^+\}^*$) and upgoing Green's function ($G^+$), and the bottom row shows the upgoing focusing function ($f_1^-$) and upgoing Green's function ($G^-$). The dashed, red line denotes the time gate. The left columns show the result of irregularly sampled data after 12 iterations of the standard scheme. The middle columns show the results when using our scheme on the same data (\autoref{is_model}), again 12 iterations were used. Finally, the 3rd column shows the reference result, obtained after 12 iterations of the standard scheme with well-sampled data. Each panel is scaled with respect to it's own maximum value.}
\label{is_results}
\end{figure}

The performance of the proposed scheme is tested on synthetic data for the model in \autoref{is_model}. For convenience, the densitiy and velocity contrasts are chosen to be the same in each layer. In total 601 sources and receivers were used with an initial spacing of 10 meters, and subsequently 50\% of the sources were randomly removed to simulate irregular sampling. The reflection response is modeled with a flat source spectrum between 5 and 80 Hz, and the direct arrival of the Green's function is modeled in a smooth version of the model. \\
We compute the results of both the proposed and standard scheme with irregularly sampled data, and a reference result of the standard scheme using well-sampled data. Each of these scenarios terminates after 12 iterations. The first column in \autoref{is_results} shows the effect of irregular sampling on the standard scheme, three different distortions are observed: sampling artifacts/distortions (e.g. top panel around 0.8 s), incorrect amplitudes (e.g. top panel at 0 to 0.4 s), and missing events (e.g. bottom panel around 0 s). While the results of the proposed scheme in the second column still show some signs of these distortions, the artifacts are largely suppressed in these results. Some edge effects in the results are introduced by the MDD. Note the resemblance between the second and third column, the latter displays the results of regular sampled data in the standard scheme. While not all artifacts are successfully removed from the data, the proposed scheme clearly matches the results of the regular data more closely than the standard scheme. 

\section{Conclusions}

The Marchenko method requires regularly sampled and collocated sources and receivers. Recently, it was shown that point-spread functions (PSFs) can, theoretically, correct for irregularly sampled sources. Here, we integrate these PSFs in the iterative Marchenko scheme, allowing for their application in more practical situations. This requires a few adaptations to the iterative scheme. Three additional steps are introduced to each iteration. First, an estimate of the inverse of the focusing function is calculated, for $f^+_1$ this inverse equals the transmission response, and for $f^-_1$ it is equal to quantity $Y$. Next, the PSF can be approximated with the aid of these inverses, the PSF describes the effects of the irregular sampling on the data. The third step is applying the PSF on the blurred responses using a MDD, resulting in regularly sampled Green's and focusing functions. Next, a time-gate can again be used to separate the focusing functions from the Green's functions. A numerical example shows clear improvement of the proposed scheme compared to the regular scheme; the results of the proposed scheme more closely resemble the regularly sampled reference. The newly proposed scheme alleviates the requirement for regularly sampled sources when using the Marchenko method. Ideally, the need for well-sampled receivers should be removed as well, this is subject to further research. By relaxing the need for perfectly sampled data, the Marchenko method is more easily applied to field data.


\section{Acknowledgements}

The authors thank Jan Thorbecke and Christian Reinicke for help with the numerical examples and insightful discussions. This research was funded by the European Research Council (ERC) under the European Union's Horizon 2020 research and innovation programme (grant agreement No: 742703).

%
%
\bibliography{Irr_bib}

\end{document}